\documentstyle[11pt,newpasp,twoside]{article}
\def\edcomment#1{\iffalse\marginpar{\raggedright\sl#1\/}\else\relax\fi}
\reversemarginpar

\begin{document}
\title{Timescales of Disk Evolution and Planet Formation}
 \author{Ray Jayawardhana}
\affil{Department of Astronomy, University of California, Berkeley, CA 94720, U.S.A.}

\begin{abstract}
It has been suggested that circumstellar disks evolve from dense, actively 
accreting structures to low-mass, replenished remnants. During this 
transition, grains may assemble into planetesimals, or the disk may be 
cleared by newborn planets. Recently identified nearby groups of young stars 
provide valuable laboratories for probing disk evolution. I discuss the 
properties of dust disks in the TW Hydrae Association and the MBM 12 cloud, 
and compare the results to other studies of disk evolution and planet
formation timescales. 
\end{abstract}

\vspace*{-0.8cm}
\section{Introduction}
Planetary systems form out of circumstellar disks that are the remnants 
of star formation (Shu, Adams, \& Lizano 1987). The young T Tauri stars with 
ages of $\sim$1 Myr, which frequently have optically-thick, 
actively-accreting disks (Strom, Edwards \& Skrutskie 1993), are thought to 
represent a stage prior to the main epoch of planet formation. In contrast, 
the dust disks of the so-called Vega-like objects, main sequence stars whose 
ages could be as large as 1 Gyr, appear to be more evolved; most of 
the dust has presumably coagulated into planets or planetesimals,
and the remaining dust in the optically-thin disk is continually replenished 
by collisions between larger bodies (Lagrange, Backman \& Artymowicz 2000). 
However, the age at which this transition occurs is not well constrained. 
One reason is a paucity of intermediate-age ($t \sim$10 Myr) nearby
stars in current samples. Unfortunately, the age estimates of early-type 
isolated main sequence stars are highly uncertain.

The two competing theories for making gas giant planets in disks predict very 
different timescales. Core accretion mechanism, in which a solid core builds 
up from icy and rocky planetesimals and later acquires an envelope of nebular 
gas, is expected to take $\sim$10 Myr (e.g., Pollack et al. 1996). 
On the other hand, the disk instability mechanism, where a gravitationally 
unstable disk fragments directly into self-gravitating protoplanetary clumps, 
is a much faster process, forming giant planets in $\sim$0.1 Myr (e.g.,
Boss 2000). Therefore, better observational constraints on disk evolution 
timescales may allow us to distinguish between the two scenarios.

\section{Insights from Nearby Stellar Groups}
The ROSAT all-sky X-ray survey, combined with follow-up spectroscopy and
proper motion data, has led to the recent identification of several groups
of young stars within 100 pc of the sun. The proximity and age differences 
of these groups make them suitable laboratories for detailed studies of disk 
evolution and planet formation. 

The TW Hydrae Association (TWA), at a distance of $\sim$55 pc and not 
associated with any obvious parent cloud, is of particular interest because 
of its $\sim$10-Myr age, as estimated using a variety of techniques. 
Our mid-infrared observations, carried out using the 
OSCIR instrument on Keck and the Cerro Tololo 4-meter telescope over the past 
two years, show that many of the TWA stars have little or no disk emission 
at 10$\mu$m (Jayawardhana et al. 1999b). Even among the five stellar systems 
with 10$\mu$m excesses, most show some evidence of inner disk evolution. The 
disk around the A0V star HR 4796A has an $r \approx 50$ AU central hole in 
mid-infrared images (Jayawardhana et al. 1998; Koerner et al. 1998). The SEDs 
of HD 98800 and Hen 3-600A also suggest possible inner disk holes 
(Jayawardhana et al. 1999a). The modest excess we detected from 
CD -33$^{\circ}$7795 could well be due to a faint companion. Only TW Hya itself
appears to harbor an optically thick, actively accreting disk of the kind 
observed in $\sim$1-Myr-old classical T Tauri stars; it is the only one with 
a large H$\alpha$ equivalent width (-220 \AA). 

If the TWA stars are indeed 10 million years old, our results suggest that 
their inner disks have already depleted either through coagulation of dust 
or accretion on to the central star. The fact that only one (TW Hya) out of 
a sample of 16 TWA members shows classical T Tauri characteristics --compared 
to $\sim$50\% --90\% of $\sim$1-Myr-old stars in star-forming regions-- argues 
for relatively rapid evolution of inner disks in pre-main-sequence (PMS) stars.
Observations at far-infrared and sub-millimeter wavelengths may reveal 
whether most TWA stars still retain their outer disks. 

We have also obtained mid-infrared photometry and high-resolution optical 
spectra for eight confirmed members of the MBM 12 high-latitude cloud at 
$\sim$65 pc. These stars are estimated to be 1-3 Myr in age. Interestingly, 
in contrast to the TWA sample, the majority of MBM 12 stars do show signatures
of optically thick, actively accreting disks (Jayawardhana et al. 2000). 

\section{Comparison to Other Studies}
Our results for TWA and MBM 12 are consistent with those of Skrutskie et al. 
(1990) who studied infrared excess in a sample of ``solar-type'' 
PMS stars in Taurus-Auriga. Of stars estimated to be younger 
than 3 Myr, Skrutskie et al. found that roughly half have optically thick 
disks extending close to the stellar surface whereas fewer than 10\% of 
somewhat older stars showed evidence for similar disks. Therefore, the 
authors suggested that few disks survive beyond 10 Myr. 

More recently, Brandner et al. (2000) have compared disk properties between
Chamaeleon and Sco-Cen star-forming regions to derive disk evolution 
timescales. They find that the 5-15 Myr-old stars in Sco-Cen have a 
15$\mu$m/7$\mu$m spectral index intermediate between that of younger 
(1-5 Myr-old) Chamaeleon objects and that of pure stellar photospheres.
Brandner et al. suggest that the difference is due to a deficiency of small 
dust grains relative to large dust grains in the Sco-Cen disks. If it is
an evolutionary effect, dust depletion of disks must start at an age between
5 and 15 Myr, again in agreement with our findings for the TWA and MBM 12 
groups. Meteoritic evidence also suggests that planetary material was made 
over a period of $\sim$ 10 Myr in the solar nebula (Podosek \& Cassen 1994; 
Wadhwa \& Russell 2000). 

On the other hand, in high-mass star-forming regions such as Orion, disks
may dissipate much faster (in $<$ 1 Myr) due to irradiation from neighboring 
luminous stars (e.g., Bally et al. 1998). It is not clear whether planets
can form in such rapidly photoevaporating disks.

\section{Future Prospects}
It is important to study a sufficiently large sample of stars with a range
of ages and environments to place strong constraints on disk evolution and 
planet formation
timescales. The identification of additional groups of young stars in the 
solar neighborhood will help, as will the unprecedented sensitivity of 
SIRTF and SOFIA to detect even optically thin disk emission from clusters at 
much greater distances than hitherto possible. There may not be a universal
timescale for disk evolution, especially when the effects of binary companions
and external irradiation are taken into account. More studies are also needed
to better understand the dissipation of the gaseous component of disks and to
determine whether it persists long enough to allow for slow accumulation 
of gas onto giant protoplanets, as predicted by the core accretion scenario.

\end{document}